\documentclass[fleqn,12pt,twoside]{article}
\usepackage{espcrc1}

\usepackage{graphicx}
\usepackage{epsfig}
\usepackage[figuresright]{rotating}

\newcommand{\AmS}{{\protect\the\textfont2
  A\kern-.1667em\lower.5ex\hbox{M}\kern-.125emS}}

\usepackage{amssymb}

\usepackage{amsmath}

\title{Quark mass dependence of nucleon properties and \\  extrapolations from Lattice QCD\thanks{Presented by W. Weise at the International School ``Quarks in Hadrons and Nuclei'', Erice, Sicily, September 2002. Work supported in part by BMBF and DFG. Preprint No. TUM-T39-02-31}}                                                                                                                                                                                                                                                       

\author{T.R. Hemmert\address[TUM]{Physik-Department, Technische Universit\"at M\"unchen, 85747 Garching, Germany},
        M. Procura\addressmark[TUM]\address[ECT]{ECT*, Villa Tambosi, 38050 Villazzano (Trento), Italy} and
        W. Weise\addressmark[TUM]\addressmark[ECT]}

\begin{document}
%
\maketitle

\begin{abstract}
We summarize developments concerning the quark mass dependence of nucleon magnetic moments and the axial-vector coupling constant $g_A$. The aim is to explore the feasibility of chiral effective field theory methods for the extrapolation of lattice QCD results, from the relatively large quark masses that can be handled in such computations down to the physically relevant range.
\end{abstract}
%
%
%

\section{Introduction}
%
QCD at low energy is realized in the form of an effective field theory with spontaneously broken chiral symmetry. Its degrees of freedom are massless Goldstone bosons (pions) interacting with Fermion sources (baryons). Chiral symmetry is broken explicitly by the non-zero quark masses. The explicit symmetry breaking is responsible for giving the pions their observed mass.\\
The spontaneous and explicit chiral symmetry breaking pattern is encoded in the relation \cite{E1}
\begin{equation}
m_\pi^2 f_\pi^2=- \frac{1}{2}(m_u+m_d)\,<\bar{u}u+\bar{d}d>+\, {\cal{O}}(m_q^2) \label{eq1}
\end{equation}
which connects the pion mass $m_\pi \simeq 0.14$ GeV and the pion decay constant $f_\pi=92.4$ MeV  with the quark mass and the chiral (quark) condensate $<\bar{q} q>$. Eq.(\ref{eq1}) displays the leading term in the ``strong condensate scenario'', with a large vacuum expectation value of the scalar quark densities, $<\bar{u}u>\,=\,<\bar{d}d>\, \simeq\, -(250\, {\rm MeV})^3$. This scenario has recently been confirmed, at least for two-flavor QCD, by a detailed chiral perturbation theory analysis \cite{E2} of improved $\pi \pi$ scattering data. Eq.(\ref{eq1}) can thus serve as a translation between the quark mass and the pion mass.\\

Lattice QCD is becoming a powerful tool for studying the structure of the nucleon \cite{E3,E4}. In practice, however, these computations are so far limited to relatively large quark masses. The ``light'' quark masses that can be handled on the lattice are typically an order of magnitude larger than the small $u$- and $d$-quark masses, $m_{u,d} < 10$ MeV, determined at a renormalization scale around 1 GeV.\\

Given this situation, methods of interpolation between lattice results and actual observables determined at the physical pion mass have become an important issue of great interest. Chiral effective field theory can in principle provide such extrapolations. First steps in this direction were taken by the Adelaide group \cite{E5} who used Pad\'e approximants based on the leading $m_\pi$-dependences as dictated by chiral symmetry.\\

This report summarizes results \cite{E6,E7} of a recent systematic analysis of the nucleon magnetic moments, $\mu_p$ and $\mu_n$, and the axial-vector coupling constant $g_A$. This analysis is based on chiral effective field theory (ChEFT). We compare two approaches: truncated ChEFT with only pions and nucleons as active degrees of freedom and a version with explicit inclusion of the $\Delta(1232)$. It turns out that in order to approach the physical values of $\mu_{p,n}$ and $g_A$, explicit $\Delta$ degrees of freedom are crucial. 

\section{Effective Field Theory}
\subsection{Framework}
%

We are working within the framework of chiral perturbation theory (ChPT) \cite{E8,E9}. It represents a systematic expansion of observables in terms of low-momentum and small quark mass scales. Nucleons are introduced as heavy Fermion fields which act as pion sources. We use the non-relativistic expansion of the theory (Heavy-Baryon Chiral Perturbation Theory, HBChPT) in orders of $1/M$ where the nucleon mass $M$ is treated as a large scale.\\

The $\Delta$(1232) isobar is the lowest spin $3/2$ excitation of the nucleon, reached by a strong magnetic ($J^P=1^-$) or axial dipole ($J^P=1^+$) transition. It must therefore be an important ingredient in considerations of the nucleon's magnetic and axial structure. Its mass differs from that of the nucleon by less than $0.3$ GeV. This is small compared to the spontaneous chiral symmetry breaking scale $4\pi f_\pi \sim 1$ GeV. Incorporating the $\Delta$ as an explicit degree of freedom is therefore mandatory in the present context of $m_\pi$ larger than $0.3$ GeV. In standard, truncated ChPT, effects of the $\Delta$(1232) are commonly relegated to some non-leading order instead.\\ 

The procedure of including spin $3/2$ baryons in the chiral effective theory is not unique. It requires accomodating the additional small scale $\Delta=M_{\Delta}-M_N \simeq 0.3$ GeV with the traditional power counting in small momenta and quark masses. We follow here the scheme proposed in \cite{E10}. In addition, the leading off-diagonal magnetic $N-\Delta$ transition is promoted to leading order as described in detail in ref. \cite{E6}. This is an essential step not only in view of the strength of the magnetic $N \to \Delta$ transition operator, but also because this operator generates non-analytic quark mass dependence which plays an important role in the extrapolation of nucleon magnetic moments. For the axial sector the standard counting scheme of ChPT appears to be sufficient in order to capture the dominant quark mass dependence. We use the scheme developed in \cite{E11} where leading axial-vector current effects related to the $\Delta$(1232) are already accounted for.  

\subsection{Effective Lagrangian}
The effective Lagrangian used for our complete one-loop calculation of $\mu_{p,n}$ and $g_A$ is
\begin{equation}
{\cal L}={\cal L}_{N}+{\cal L}_{N\Delta}+{\cal L}_{\Delta}+{\cal L}_{\pi\pi}\,,
\end{equation}
where
\begin{equation}
{\cal L}_{\pi\pi}=\frac{f_\pi ^2}{4} \, \mathrm{Tr}\, [\nabla_{\mu}U^{\dag} \nabla^{\mu}U]+ \mathrm{mass\;term\,.}
\end{equation}

The chiral field is represented as 
\begin{equation}
U=\left(1-\frac{\pi_a^2}{f_\pi^2}\right)^{1/2}+\frac{i}{f_\pi}\tau_a \pi^a
\end{equation}
with Goldstone boson fields $\pi_a$ and standard $SU(2)$ isospin matrices $\tau_a$. The (gauge) covariant derivative $\nabla^{\mu}$ includes the coupling to the electromagnetic field. The leading order $\pi N N$, $\pi \Delta \Delta$ and $\pi N \Delta$ Lagrangians are of the generic form:
\begin{equation}
{\cal L}^{(1)}_{N}=\bar{N}_v\left[i\,v\cdot D+g_A\,S\cdot u\right]N_v
\label{eq5}
\end{equation}
\begin{equation}
{\cal L}^{(1)}_{\Delta}=\,\bar{T}_i^\mu\left[(\delta^{ij}-\frac{1}{3}\tau^i \tau^j)\Delta -iv\cdot D^{ij}-g_1\,S\cdot u\,\delta^{ij}\right]g_{\mu\nu}\,T_j^\nu\; \label{eq6}
\end{equation}
\begin{equation}
{\cal L}^{(1)}_{N\Delta}=\bar{T}^\mu_i\,\mathrm{Tr}\left[\frac{{\tau}^i}{2}(c_A u_\mu+c_V\,i\,f_{\mu\nu}\,S^{\nu})\right]\,N_v\,+\,h.c.\; ,\label{eq7}
\end{equation}
where $N_v$ and $T^{\mu}$ are non-relativistic spin $1/2$ and spin $3/2$ fields, $v^{\mu}$ is the velocity and $S^{\mu}$ the spin; the $D^{\mu}$ are the covariant derivatives which incorporate the coupling to Goldstone bosons as well as electromagnetic fields, and $u^{\mu}$ is an axial-vector combination of chiral fields and their covariant derivatives. Additional isospin structure is indicated by the $(ij)$ indices in ${\cal L}^{(1)}_{\Delta}$. The mass term in eq.(\ref{eq6}) involves the $N$-$\Delta$ mass difference $\Delta=M_{\Delta}-M_N$ multiplied by the isospin $I=3/2$ projector. The $N$-$\Delta$ transition Lagrangian (\ref{eq7}) includes an axial term (with coupling constant $c_A$) and a vector term (with coupling constant $c_V$). The latter one couples to the electromagnetic field through a tensor $f_{\mu \nu}$. The detailed expressions for all leading order terms (4-6) are given in ref.\cite{E6}.\\

Up to this point there are four constants representing the intrinsic structure of the interacting system of pions, nucleons and $\Delta$'s in the presence of an external electromagnetic field: the axial-vector coupling constant (actually to be taken in the chiral limit, $m_\pi \to 0$, and then denoted by $g_A^0$), with its empirical value $g_A=1.267$; the axial coupling $g_1$ of the $\Delta$(1232) which is treated as a parameter; and the leading vector and axial $N$-$\Delta$ transition couplings $c_{V,A}$ just mentioned.\\

A consistent one-loop calculation requires next-to-next-to-leading order terms in the effective Lagrangian. Their detailed forms are again given in refs.\cite{E6,E7} and will not be repeated here. The corresponding ${\cal L}^{(2)}_{N}$ involves the anomalous isovector and isoscalar magnetic moments, $\kappa_S^0$ and $\kappa_V^0$, all taken in the chiral limit. Renormalization at the one-loop level introduces additional counterterms encoding short-distance dynamics, with corresponding constants that need to be determined from suitable experimental data. It turns out that the number of additional parameters of this sort at next-to-leading order is quite limited: there is only one such parameter for either one of the magnetic moments or for the determination of $g_A$.

\section{Magnetic moments}
\begin{figure}[!htb]
  \begin{center}
    \includegraphics*[width=0.6\textwidth,height=10cm]{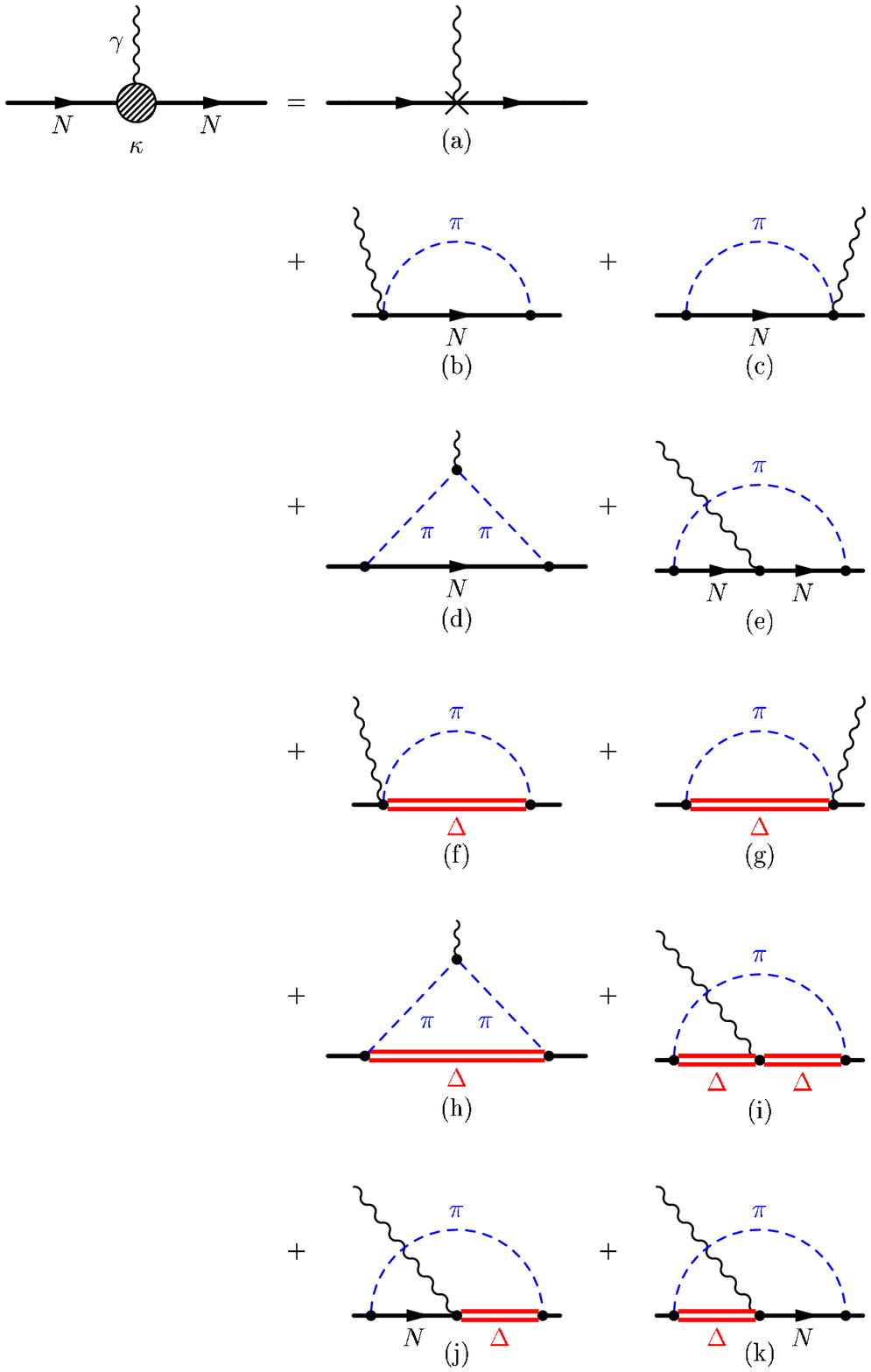}
     \caption{Leading-one-loop diagrams included in the chiral effective field theory approach to nucleon magnetic moments. Diagrams (a)-(e) are the ones incorporated in the truncated heavy-baryon chiral perturbation theory. Diagrams (f)-(k) involve explicit $\Delta$(1232) degrees of freedom.}
\label{fig1}
  \end{center}
\end{figure}

We start with the calculation of the isovector anomalous magnetic moment of the nucleon, $\kappa=\mu_p - \mu_n -1$. To leading one-loop order, the relevant set of diagrams is shown in Fig.\ref{fig1}, with the electromagnetic field as an external source. The result of the calculation \cite{E6} can be expressed in closed form:
\begin{equation}
\kappa = \kappa_V^0 -\frac{g_A^2 M}{4\pi f_\pi^2} m_\pi + F(m_\pi, \Delta)\,. \label{eqk}
\end{equation}
The function $F$ is given in detail in ref.\cite{E6}. It includes powers and logarithms of the dimensionless quantity $m_\pi / \Delta$ with non-analytic quark mass dependence. It is instructive to expand $F$ around the chiral limit. To order $m_\pi^3$ one finds:
\begin{equation}
F(m_\pi,\Delta)= m_\pi^2 \frac{c_A^2}{9 \pi^2 f_\pi^2} \left[\frac{M}{\Delta} \left(\ln{\frac{m_\pi}{2 \Delta}} - \frac{1}{2} \right) + \frac{2}{3} \frac{c_V}{c_A} g_A M \left(1+\frac{2 m_\pi}{\Delta}\right)+C_1\right]\,.
\label{eq8}
\end{equation}

\begin{figure}[!htb]
  \begin{center}
    \includegraphics*[width=0.7\textwidth,height=8cm]{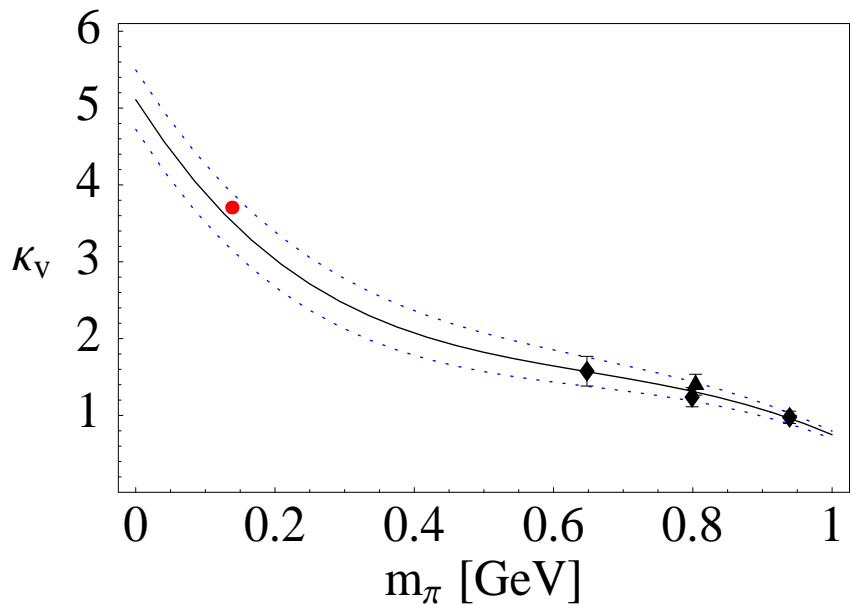}
     \caption{Pion mass dependence of the isovector anomalous magnetic moment as compared to lattice data taken from ref.\cite{E5}.}
\label{fig2}
  \end{center}
\end{figure}

The adjustable constant $C_1$ combines a logarithm and the finite part of a counterterm. The expansion (\ref{eq8}) holds at small pion masses. The actual calculations are performed with the full expression for $F(m_\pi,\Delta)$ as explained in ref.\cite{E6}. We can now fix the three constants that remain undetermined by other observables, $\kappa_V^0$, $c_V$ and $C_1$, to three lattice data for $\kappa_V$ corresponding to the lowest possible values of the quark (pion) mass, and extrapolate down to the physically relevant value of $m_\pi$ and further down to the chiral limit. The result is shown in Fig.\ref{fig2}. It is indeed remarkable and highly non-trivial that this extrapolation meets the empirical $\kappa_V \simeq 3.71$ (in units of nuclear magnetons) within the uncertainties given by the limited accuracy of the lattice data. The resulting isovector anomalous magnetic moment in the chiral limit is $\kappa_V^0 \simeq 5.1$, and the vector coupling of the $\Delta$ is $c_V \simeq 2.3\, {\mathrm{GeV}}^{-1}$ from this approach. All other constants have been kept at their physical values ($g_A=1.267$, $c_A=1.125$, $f_\pi=92.4\, \mathrm{MeV}$, $M=0.939\, \mathrm{GeV}$ and $\Delta \equiv M_{\Delta}-M=0.271\, \mathrm{GeV}$). Note that for this scheme to work, the role of the $\Delta$(1232) as an explicit degree of freedom is crucial. Truncated chiral perturbation theory with pions and nucleons only (i.e. with $F=0$ in NLO, eq.(\ref{eqk})) would obviously not work at all at this order.\\

The lattice data used here for reference are computed in the quenched approximation. However, at relatively large quark masses where those data have been taken, effects of ``unquenching'' (i.e. quark loop effects) turn out to be within the errors of the quenched calculation \cite{E15}. It is therefore legitimate to use the ``full'' (unquenched) effective field theory in order to extrapolate down to small quark masses.

\begin{figure}[!htb]
  \begin{center}
    \includegraphics*[width=0.7\textwidth, height=8cm]{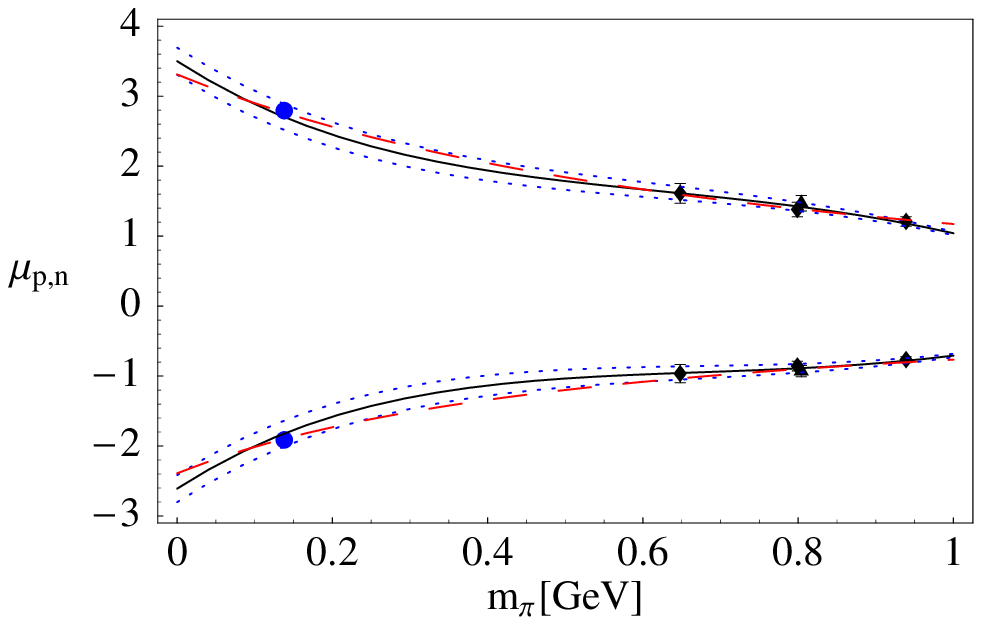}
    \caption{Pion mass dependence of proton and neutron magnetic moments as computed in ref.\cite{E6} (solid), compared to lattice data and a Pad\'e-fit formula (dashed) reported in ref.\cite{E5}. The empirical $\mu_p=2.793$ and $\mu_n=-1.913$ are also shown.}
\label{fig3}
  \end{center}
\end{figure}

The small isoscalar anomalous magnetic moment, $\kappa_S=\mu_p + \mu_n -1$ taken to the same leading one-loop order, is analytic in the quark mass: $\kappa_S= \kappa_S^0 - \mathrm{const.}\cdot m_\pi^2$, where the constant parametrizes short-distance physics. Combining isovector and isoscalar results one obtains the quark (pion) mass dependence of the proton and neutron magnetic moments \cite{E6} as shown by the full lines in Fig.\ref{fig3}, again compared to lattice data. Although these data correspond to pion masses larger than $0.6$ GeV, one achieves an extrapolation to the physical region which comes remarkably close to the empirical magnetic moments - a non-trivial result. Within the error band given, this result turns out also to be surprisingly close to the Pad\'e-fit extrapolation of ref.\cite{E5}.

\section{Axial-vector coupling constant}

Next, we turn to the axial-vector coupling constant $g_A$ of the nucleon. In a leading-one-loop calculation, the external field in Fig.\ref{fig1} is replaced by an axial-vector source. Diagrams (d) and (h) do not exist in this case, whereas diagrams (f) and (g) can only start contributing at next-to-leading-one-loop order; instead at leading-one-loop there is one extra graph involving an axial $\pi \pi N N$ four-point vertex with a pion loop attached.

The complete one-loop expression incorporating nucleon and $\Delta$(1232) has the form

\begin{eqnarray}
g_A=g_A^0-\frac{{g_A^0}^3}{(4\pi f_\pi)^2}m_\pi^2+G(m_\pi,\Delta),
\end{eqnarray}
where $g_A^0$ is the axial-vector coupling constant in the chiral limit. The function $G(m_\pi, \Delta)$, specified in detail in ref.\cite{E7}, is non-analytic in the quark mass and includes logarithms in $m_\pi/\Delta$. The expansion of $G$ around the chiral limit gives

\begin{eqnarray}
G\simeq m_\pi^2 \left[C_2+\frac{c_A^2}{4\pi^2 f_\pi^2}\left(\frac{155}{972}g_1-\frac{17}{36}g_A^0 \right)\right] + \, {\cal{O}}(m_\pi^2 \ln{m_\pi}) + \, {\cal{O}}(\frac{m_\pi^3}{\Delta})\,.
\end{eqnarray}
 It involves the axial couplings $g_1$ of the $\Delta$(1232) and $c_A$ for the $N \Delta$ transition, see eqs.(\ref{eq5}, \ref{eq6}). It also involves a parameter $C_2$ which summarizes short-distance information that enters in higher-order contact terms. It might then appear on first sight that the number of poorly known input parameters ($g_A^0$, $g_1$ and $C_2$) is prohibitively large. However, $C_2$ is actually constrained by an independent ChPT analysis of $\pi N \to \pi \pi N$ processes \cite{E12}. Such processes include vertices in which a nucleon interacts with three pions, and these contribute in turn to the axial form factor of the nucleon when combined to spin and parity $1^{+}$ in the $t$-channel. Such connections are in fact characteristic of the effective field theory approach which establishes detailed relationships between seemingly independent observables.

\begin{figure}[!htb]
  \begin{center}
    \includegraphics*[width=0.65\textwidth,height=7cm]{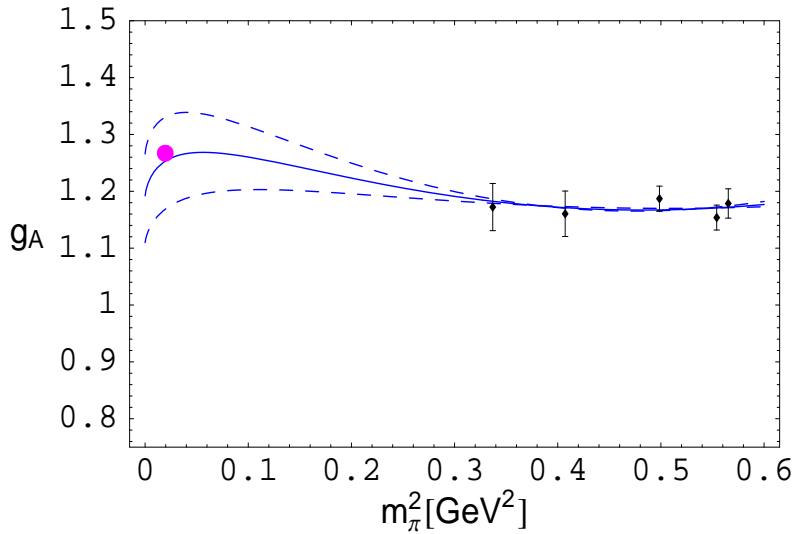}
    \caption{Chiral extrapolation of $g_A$ \cite{E7} from lattice data \cite{E4}. The dashed curves give an impression of the uncertainty range resulting from the input obtained from the $\pi N \to \pi \pi N$ analysis. The empirical value $g_A=1.267$ is also shown.}\label{fig4}
  \end{center}
\end{figure}

The expansion (\ref{eq8}) holds for small pion masses. The actual calculations of the quark mass extrapolation of $g_A$ have again been performed with the full NLO expression for $G(m_\pi, \Delta)$ at the one-loop level. Once the counterterm entering $G$ is constrained by the $\pi N \to \pi \pi N$ analysis, two remaining parameters ($g_A^0$ and $g_1$) can be fixed by the (quenched) lattice data of the QCDSF-UKQCD collaboration \cite{E4}. The extrapolation down to the physical region works astonishingly well, as demonstrated in Fig.\ref{fig4}. The error band surrounding the solid curve in this figure reflects primarily uncertainties in the constraint on $C_2$. Once again the contributions involving the $\Delta$(1232), contained in the function $G$, are crucial in achieving this result. Truncated ChPT with only $\pi$ and $N$ degrees of freedom fails badly at this point.

Extrapolations of the moments of spin structure functions recently investigated by the Adelaide group \cite{E13} come to similar conclusions concerning the importance of the $\Delta$(1232) in spin observables of the nucleon but do not show the upturn in $g_A$ at low $m_\pi^2$.

The importance of the explicit $\Delta$(1232) degrees of freedom in the discussion of $g_A$ does not come as a surprise. Recall the Adler-Weisberger sum rule \cite{E14},
\begin{eqnarray}
g_A^2&=&1+\frac{2f_\pi^2}{\pi}\int_{m_\pi}^{\infty}\frac{d\omega}{\sqrt{\omega^2-m_\pi^2}}[\sigma_{\pi^+ p}(\omega)-\sigma_{\pi^- p}(\omega)]+{\cal{O}}\left(\frac{m_\pi^2}{M_N^2}\right)\;.
\end{eqnarray}
It relates the surplus of $g_A$ beyond its trivial value $g_A=1$ (for a pointlike, structureless nucleon) to the excess of the $\pi^{+}p$ over the $\pi^{-}p$ cross section, a feature dominated by $\Delta$(1232) resonance excitation.

\section{Conclusions and outlook}

We have demonstrated the feasibility of systematic chiral extrapolations of nucleon properties from lattice QCD, down to the range of realistic light-quark masses where the comparisons with the actual onservables can be made. An important element in this approach is the treatment of the $\Delta$(1232) isobar as an \emph{explicit} degree of freedom in the underlying chiral effective field theory. The $\Delta$ is in fact the prime agent which governs important spin-flip dynamics, through the $N \Delta$ transition, at the level of the chiral effective theory of pions coupled to baryons representing low-energy QCD.

Extrapolations of the kind presented here are still subject to uncertainties because the lowest pion masses that can presently be handled on the lattice ($m_\pi \gtrsim 0.5$ GeV) may still be too large. The fact that these extrapolations work using a well organized scheme based on chiral effective field theory is nevertheless encouraging. Once lattice calculations reach the $0.3$ GeV frontier for the pion mass, the remaining gap between lattice results and actual observables may well be closed reliably by analytical methods as presented here.

Discussions with W. Melnitchouk and A.W. Thomas during the preparation of this manuscript are gratefully appreciated.  

\newpage




\begin{thebibliography}{99}
\itemsep=0cm
%
\bibitem{E1}M. Gell-Mann, R. Oakes and B. Renner, Phys. Rev. {\bf 122} (1968) 2195.
%
%
\bibitem{E2}G. Colangelo, J. Gasser and H. Leutwyler, Phys. Rev. Lett. {\bf 86} (2001) 5008.
%
%
\bibitem{E3}J. Negele, Nucl. Phys. {\bf A699} (2002) 18, and references therein.
%
%
\bibitem{E4}G. Schierholz, Proc. Int. Conf. BARYONS 2002 (World Sci., Singapore), to appear, and private communication.
%
%
\bibitem{E5}D.B. Leinweber, D.H. Lu and A.W. Thomas, Phys. Rev. {\bf D60} (1999) 034014.
%
%
\bibitem{E6}T.R. Hemmert and W. Weise, Eur. Phys. J. {\bf A15} (2002) 487.
%
%
\bibitem{E7}T.R. Hemmert, M. Procura and W. Weise, preprint TUM-T39-02-22 (2002).
%
%
\bibitem{E8}S. Weinberg, Physica {\bf A96} (1979) 327;\\J. Gasser and H. Leutwyler, Ann. of Phys. {\bf 158} (1984) 142.
%
%
\bibitem{E9}see: A.W. Thomas and W. Weise, The Structure of the Nucleon, Wiley-VCH, Berlin, 2001, and refs. therein.
%
%
\bibitem{E10}T.R. Hemmert, B. Holstein and J. Kambor, Phys. Lett. {\bf B395} (1997) 89; J.Phys {\bf G24} (1998) 1831.
%
%
\bibitem{E11}V. Bernard, H.W. Fearing, T.R. Hemmert and U.-G. Meissner, Nucl. Phys. {\bf A635} (1998) 121; ibid. {\bf A642} (1998) 563.
%
%
\bibitem{E12}N. Fettes, V. Bernard and U.-G. Meissner, Nucl. Phys. {\bf A669} (2000) 269.
%
%
\bibitem{E13}W. Detmold, W. Melnitchouk and A.W. Thomas, Phys. Rev. {\bf D66} (2002) 054501.
%
%
\bibitem{E14}S.L. Adler, Phys. Lett. {\bf 14} (1965) 1051;\\ W.I. Weisberger, Phys. Lett. {\bf 14} (1965) 1047.
%
%
%
\bibitem{E15}D.B. Leinweber et al., private communication and archive no. {\tt hep-lat/0211017}.
%

\end{thebibliography}
\end{document}